\documentstyle[12pt]{article}
 \makeatletter
 \def\leqq{\mathrel{\mathpalette\gl@align<}}
 \def\geqq{\mathrel{\mathpalette\gl@align>}}
 \def\gl@align#1#2{\lower.6ex\vbox{\baselineskip\z@skip\lineskip\z@
     \ialign{$\m@th#1\hfil##\hfil$\crcr#2\crcr=\crcr}}}
 \makeatother

 \makeatletter
 \def\sileqq{\mathrel{\mathpalette\gs@align<}}
 \def\sigeqq{\mathrel{\mathpalette\gs@align>}}
 \def\gs@align#1#2{\lower.6ex\vbox{\baselineskip\z@skip\lineskip\z@
     \ialign{$\m@th#1\hfil##\hfil$\crcr#2\crcr\sim\crcr}}}
 \makeatother
\begin{document}
\hbadness=10000
\hbadness=10000
\begin{titlepage}
\nopagebreak
\begin{flushright}
\end{flushright}
\vspace{1.5cm}
\begin{center}

{\large \bf Cubic Matrix, Nambu Mechanics and Beyond}

\vspace{1cm}
 
{Yoshiharu Kawamura}\footnote{E-mail:
haru@azusa.shinshu-u.ac.jp} 

\vspace{0.7cm}
Department of Physics, Shinshu University,
Matsumoto 390-8621, Japan\\
\end{center}
\vspace{0.7cm}

\nopagebreak

\begin{abstract}
We propose a generalization of cubic matrix mechanics by introducing a canonical triplet
and study its relation to Nambu mechanics.
The generalized cubic matrix mechanics we consider can be interpreted 
as a $\lq$quantum' generalization of Nambu mechanics.
\end{abstract}
\vfill
\end{titlepage}
\pagestyle{plain}
\newpage
\def\thefootnote{\fnsymbol{footnote}}

\section{Introduction}

The study of a new, generalized mechanics beyond classical mechanics (CM) and quantum mechanics (QM)
is often regarded as ambitious, because QM has been applied to very broad
areas of physics with indisputable success.
There is, however, no strong reason to believe
that QM is the unique mechanics to describe nature at a fundamental level 
(around and beyond the gravitational scale).
In fact, M-theory \cite{M} is a promising candidate of a fundamental theory of nature,
and there is an intriguing proposal for a formulation of M-theory 
based on the infinite momentum frame\cite{M2}, deeply related to
the quantum mechanics of supermembranes\cite{mem},
but no complete such formulation has yet been made.
There is a possibility that an ultimate theory requires a new mechanics
combined with a configuration of fundamental objects.
Therefore it is still meaningful to construct a new, generalized mechanics and study its properties.

Nambu proposed a generalization of Hamiltonian dynamics through the extension
of phase space based on the Liouville theorem and gave a suggestion
for its quantization.\cite{Nambu}
The structure of this mechanics has been studied in the framework of constrained
systems \cite{con} and in geometric and algebraic formulations.\cite{geo}
There are several works in which the quantization of Nambu mechanics (NM)
is investigated.\cite{geo,q1,q2,q3,q4,q5}
As an interesting approach, Awata, Li, Minic and Yoneya introduced many-index objects
to realize the quantum version of Nambu bracket.\cite{q3}

Recently, a new mechanics has been proposed 
based on many-index objects \cite{YK}, which is a generalization of Heisenberg's matrix mechanics,
and its basic structure has been studied from the algebraic point of view\cite{YK2}.
The definition of the triple product among three-index objects called $\lq$cubic matrices' 
given in Ref. \cite{YK}
is different from that given in Ref. \cite{q3} 
in the respect that a generalization of the Ritz rule
in the phase factor is required, but the associativity of the products is not necessary.
This mechanics possesses a counterpart to the canonical structure in CM and can be generalized through the
extension of phase space modeling, following NM.
It is quite interesting to investigate this type of generalization and its relation to NM.

In this paper, we propose a generalization of cubic matrix mechanics, which
we refer to as $\lq$generalized cubic matrix mechanics', by introducing a canonical triplet
and study the correspondence to NM.
A conjecture concernig operator formalism is also given.

Our strategy is almost the same as that in Ref. \cite{YK2}. 
In the next section, we review the canonical structure of NM
and discuss the basic structure that a mechanics beyond NM should possess.
We formulate a generalized cubic matrix mechanics and study
its structure from an algebraic viewpoint and its relation to NM in $\S$3. 
Section 4 is devoted to conclusions and discussion.

\section{Nambu mechanics and beyond}

\subsection{Canonical structure of Nambu mechanics}

Here we review the canonical structure of Nambu mechanics.\cite{Nambu}
For simplicity, we treat a system with a 3-dimensional phase space
whose variables are $x = x(t)$, $y = y(t)$ and $z = z(t)$.
They satisfy the $\lq\lq$Hamilton's equations"
\begin{eqnarray}
{d x \over dt} = {\partial (K, H) \over \partial (y, z)}, ~~
{d y \over dt} = {\partial (K, H) \over \partial (z, x)}, ~~
{d z \over dt} = {\partial (K, H) \over \partial (x, y)},
\label{Hamilton-eq}
\end{eqnarray}
where the right-hand sides represent 2-dimensional Jacobians, and $K$ and $H$ are the $\lq\lq$Hamiltonians".
Physical variables are given by functions of the canonical variables and the time variable $t$; e.g., 
$A = A(x, y, z, t)$, $B = B(x, y, z, t)$, and $C = C(x, y, z, t)$.
Hereafter we consider systems such that physical variables do not contain $t$ explicitly,
that is, closed physical systems.
The Nambu bracket of three variables $A$, $B$ and $C$ with respect to $x$, $y$ and $z$ 
is defined by
\begin{eqnarray}
\{A, B, C\}_{\scriptsize{\mbox{NB}}} 
\equiv {\partial (A, B, C) \over \partial (x, y, z)} ,
\label{NB}
\end{eqnarray}
where the right-hand side represents a 3-dimensional Jacobian.
Hence, the Nambu brackets of the canonical variables are given by
\begin{eqnarray}
&~& \{x, y, z\}_{\scriptsize{\mbox{NB}}} = 1 , 
~ \{x, x, z\}_{\scriptsize{\mbox{NB}}} = \{x, y, y\}_{\scriptsize{\mbox{NB}}} = \cdots 
= \{z, z, z\}_{\scriptsize{\mbox{NB}}} = 0 .
\label{NB-xxy}
\end{eqnarray}
The basic features of the Nambu bracket are as follows:
\begin{eqnarray}
&~& \{A, B, C\}_{\scriptsize{\mbox{NB}}} =  \{B, C, A\}_{\scriptsize{\mbox{NB}}} 
=  \{C, A, B\}_{\scriptsize{\mbox{NB}}} \nonumber \\
&~& = - \{C, B, A\}_{\scriptsize{\mbox{NB}}} = - \{B, A, C\}_{\scriptsize{\mbox{NB}}} 
= - \{A, C, B\}_{\scriptsize{\mbox{NB}}} ,
{\mbox{(skew-symmetry)}}
\label{NB-1}\\
&~& \{A + B, C, D\}_{\scriptsize{\mbox{NB}}} = \{A, C, D\}_{\scriptsize{\mbox{NB}}} 
+ \{B, C, D\}_{\scriptsize{\mbox{NB}}} , 
~~~ {\mbox{(linearity)}}
\label{NB-2}\\
&~& \{\{A, B, C\}_{\scriptsize{\mbox{NB}}}, D, E\}_{\scriptsize{\mbox{NB}}} \nonumber \\
&~& ~~~~~~~~~~~~~~~~ = \{\{A, D, E\}_{\scriptsize{\mbox{NB}}}, B, C\}_{\scriptsize{\mbox{NB}}} 
 + \{A, \{B, D, E\}_{\scriptsize{\mbox{NB}}}, C\}_{\scriptsize{\mbox{NB}}} \nonumber \\
&~& ~~~~~~~~~~~~~~~~~~~~~~ + \{A, B, \{C, D, E\}_{\scriptsize{\mbox{NB}}}\}_{\scriptsize{\mbox{NB}}} ,
~ {\mbox{(fundamental identity)}}
\label{NB-3}\\ &~& \{A B, C, D\}_{\scriptsize{\mbox{NB}}} = A \{B, C, D\}_{\scriptsize{\mbox{NB}}} 
+ \{A, C, D\}_{\scriptsize{\mbox{NB}}} B . 
~~{\mbox{(derivation rule)}}
\label{NB-4}
\end{eqnarray}
By use of (\ref{Hamilton-eq}) and (\ref{NB}), the physical variable $A$ is shown to satisfy the equation
\begin{eqnarray}
{d A \over dt} = \{A, K, H\}_{\scriptsize{\mbox{NB}}} .
\label{Hamilton-eqA}
\end{eqnarray}

We call a transformation $A \to A' = {\cal{C}}(A)$ that preserves the bracket structure 
$\lq$canonical':
\begin{eqnarray}
\{A, B, C\}_{\scriptsize{\mbox{NB}}} \longrightarrow {\cal{C}}(\{A, B, C\}_{\scriptsize{\mbox{NB}}}) 
= \{{\cal{C}}(A), {\cal{C}}(B), {\cal{C}}(C)\}_{\scriptsize{\mbox{NB}}} .
\label{canonical}
\end{eqnarray}
The infinitesimal version $A \to A' = A + \delta A$ is given by
\begin{eqnarray}
\delta A = \{A, G_1, G_2\}_{\scriptsize{\mbox{NB}}} \delta s ,
\label{inf-canonical-tr}
\end{eqnarray}
where $G_1$ and $G_2$ are generators of the transformation,
and $\delta s$ is an infinitesimal parameter.
We can show that the bracket structure is preserved under 
the transformation (\ref{inf-canonical-tr}) by using the identity (\ref{NB-3}).

\subsection{Beyond Nambu mechanics}

The structure of Nambu mechanics is so elegant 
that it is natural to expect the existence of a $\lq$quantum' counterpart of NM.
In this subsection, we present a conjecture for the basic structure of a new mechanics beyond NM
based on a requirement that {\it{the algebraic structure of equations of motion and symmetry transformations
be preserved (up to anomalous breakings)}}.

The above requirement is expressed as the following properties:
\begin{enumerate}
\item There are counterparts of the canonical variables in NM, 
which are denoted $X = X(t)$, $Y = Y(t)$ and $Z = Z(t)$, 
and physical variables are functions
of $X$, $Y$ and $Z$ in a closed system.
There exists a counterpart of the Nambu bracket, which we call the $\lq$generalized bracket', and 
the bracket relations for $X$, $Y$ and $Z$ are conditions that place restrictions on the phase space
(like quantization conditions in QM).
The generalized bracket does not necessarily possess all the algebraic properties of the Nambu bracket.
However, at least it possesses properties of skew-symmetry and linearity.

\item The equations of motion for physical variables are of the same type as those in NM.
More specifically, an equation of motion is obtained from the correspoding
equation in NM by replacing the Nambu bracket 
with the generalized bracket.

\item There is a transformation that preserves the generalized bracket structure
that we call a $\lq$generalized canonical transformation'.
The $\lq$fundamental identity' analogous to (\ref{NB-3}) for the generalized bracket 
holds, including generators,
which are conserved quantities.
Continuous symmetry transformations are realized as generalized canonical transformations
of conserved quantities.
\end{enumerate}

Now we formulate the basic structure of a new mechanics based on the above properties.
\begin{enumerate}
\item Let us denote the generalized bracket by ${\cal{B}}(*, *, *)$ and impose the following conditions 
on $X$, $Y$ and $Z$:
\begin{eqnarray}
{\cal{B}}(X, Y, Z) = \Theta , 
{\cal{B}}(X, X, Z) = {\cal{B}}(X, Y, Y) = \cdots = {\cal{B}}(Z, Z, Z) = 0 .
\label{calB-XYY}
\end{eqnarray}
Here $\Theta$ is a constant of motion, and the bracket of $\Theta$ and 
any conserved quantities $\Lambda_i$ 
vanishes; i.e. ${\cal{B}}(\Theta, \Lambda_i, \Lambda_j) = 0$.
The skew-symmetry and linearity conditions are expressed by
\begin{eqnarray}
&~& {\cal{B}}(A, B, C) = {\cal{B}}(B, C, A) = {\cal{B}}(C, A, B) \nonumber \\
&~& ~~~~~~~~~~~ = - {\cal{B}}(C, B, A) = - {\cal{B}}(B, A, C) = - {\cal{B}}(A, C, B) ,
\label{calB-1}\\
&~& {\cal{B}}(A + B, C, D) = {\cal{B}}(A, C, D) + {\cal{B}}(B, C, D) .
\label{calB-2}
\end{eqnarray}
We do not necessarily require correspondences between the fundamental identity 
nor the derivation rule in NM as properties of ${\cal{B}}(*, *, *)$ for generic variables.

\item The equation of motion for a physical quantity $A$ is given by
\begin{eqnarray}
{d A \over dt} = {\cal{B}}(A, K, H) ,
\label{eq}
\end{eqnarray}
where $K$ and $H$ are the $\lq\lq$Hamiltonians".

\item A generalized canonical transformation is defined by the transformation $A \to A' = {\cal{G}}(A)$, 
which preserves the structure of ${\cal{B}}(*, *, *)$:
\begin{eqnarray}
{\cal{B}}(A, B, C) \longrightarrow {\cal{G}}({\cal{B}}(A, B, C))
= {\cal{B}}({\cal{G}}(A), {\cal{G}}(B), {\cal{G}}(C)) .
\label{g-canonical}
\end{eqnarray}
The infinitesimal version of (\ref{g-canonical}) is written
\begin{eqnarray}
\delta {\cal{B}}(A, B, C) = {\cal{B}}(\delta A, B, C) + {\cal{B}}(A, \delta B, C) 
 + {\cal{B}}(A, B, \delta C) ,
\label{inf-g-canonical}
\end{eqnarray}
under the infinitesimal generalized canonical transformation $A \to A' = A + \delta A$.
For conserved quantities $G_1$ and $G_2$ [i.e., $d G_i/dt = {\cal{B}}(G_i, K, H) = 0$],
the fundamental identity holds:
\begin{eqnarray}
&~& {\cal{B}}({\cal{B}}(A, B, C), G_1, G_2) 
= {\cal{B}}({\cal{B}}(A, G_1, G_2), B, C) \nonumber \\
&~& ~~~~~~~~~~~ + {\cal{B}}(A, {\cal{B}}(B, G_1, G_2), C)  + {\cal{B}}(A, B, {\cal{B}}(C, G_1, G_2)) .
\label{calB-3}
\end{eqnarray}
Then, a symmetry transformation is given by 
the infinitesimal generalized canonical transformation,\footnote{
It is not certain whether every continuous generalized canonical transformation $A \to A' = {\cal{G}}(A)$
can be constructed from the infinitesimal one given by $\delta A = {\cal{B}}(A, F_1, F_2) \delta s$,
where $F_i$ are generators.
Here, we require the algebraic structure of symmetry transformations to be identical to that in NM.}
\begin{eqnarray}
\delta A = {\cal{B}}(A, G_1, G_2) \delta s .
\label{inf-g-canonical-tr}
\end{eqnarray}
\end{enumerate}


\section{Generalized cubic matrix mechanics}

We have discussed the basic structure that a new mechanics beyond Nambu mechanics should possess.
It is expected that the study of NM will be helpful
to understand the structure of M-theory\cite{M,M2}, through the quantum theory of
supermembranes\cite{mem}.
For this reason, it is important to construct a $\lq$quantum' version of NM and study its features.
In this section, we propose a new mechanics based on cubic matrices,
which is a generalization of the cubic matrix mechanics examined in Refs. \cite{YK} and \cite{YK2},
and study its structure and the correspondence to NM.

\subsection{Cubic matrix}

Here we state our definition of a cubic matrix and its related terminology.
A cubic matrix is an object with three indices, $A_{lmn}$, 
which is a generalization
of a usual matrix, such as $B_{mn}$.
We refer to a cubic matrix whose elements possess cyclic symmetry, i.e., $A_{lmn} = A_{mnl} = A_{nlm}$,
as a cyclic cubic matrix.
We define the hermiticity of a cubic matrix by 
$A_{l'm'n'}(t) = A_{lmn}^{*}(t)$ for odd permutations among indices
and refer to a cubic matrix possessing hermicity as a hermitian cubic matrix.
Here, the asterisk indicates complex conjugation.
A hermitian cubic matrix is a special type of cyclic cubic matrix, because it obeys the relations
$A_{lmn} = A_{mln}^{*} = A_{mnl} = A_{nml}^{*} = A_{nlm} = A_{lnm}^{*}$.
We refer to the following form of a cubic matrix as a normal form or a normal cubic matrix:
\begin{eqnarray}
A^{(N)}_{lmn} = \delta_{lm} a_{mn} + \delta_{mn} a_{nl} + \delta_{nl} a_{lm}  .
\label{AN} 
\end{eqnarray}
A normal cubic matrix is also a special type of cyclic cubic matrix.
The elements of a cubic matrix are treated as $c$-numbers throughout this paper.

\subsection{Generalized cubic matrix mechanics and its structure}

The physical variables are cyclic cubic matrices given by
\begin{eqnarray}
A_{lmn}(t) = A_{lmn} e^{i\Omega_{lmn}t} ,
\label{C}
\end{eqnarray}
where the angular frequency $\Omega_{lmn}$ has the properties
\begin{eqnarray}
\Omega_{l'm'n'} = \mbox{sgn}(P) \Omega_{lmn} , ~~
(\delta \Omega)_{lmnk} \equiv \Omega_{lmn} - \Omega_{lmk} + \Omega_{lnk}
 - \Omega_{mnk} = 0 .
\label{cycle}
\end{eqnarray}
Here, sgn($P$) is $+1$ and $-1$ for even and odd permutations among indices, respectively.
The operator $\delta$ is regarded as a coboundary operator 
that changes $k$-th antisymmetric objects into $(k+1)$-th objects,
and this operation is nilpotent, i.e. $\delta^2(*) =0$.\cite{cohom}
The frequency $\Omega_{lmn}$ is regarded as a $3$-cocycle, from the second equation in (\ref{cycle}).

If we define the triple product among cubic matrices $A_{lmn}(t) = A_{lmn} e^{i\Omega_{lmn}t}$, 
$B_{lmn}(t) = B_{lmn} e^{i\Omega_{lmn}t}$ and $C_{lmn}(t) = C_{lmn} e^{i\Omega_{lmn}t}$ by
\begin{eqnarray}
(A(t)B(t)C(t))_{lmn} \equiv
\sum_k A_{lmk}(t) B_{lkn}(t) C_{kmn}(t) = (A B C)_{lmn} e^{i\Omega_{lmn}t} ,
\label{cubicproduct}
\end{eqnarray}
this product takes the same form as (\ref{C}), with the relation (\ref{cycle}), which is
a generalization of the Ritz rule.\footnote{
The Ritz rule is given by $\Omega_{ln} = \Omega_{lm} + \Omega_{mn}$ in QM, 
where $\Omega_{mn}$ is the angular frequency of radiation from an atom.}
We comment that the resultant three-index object, $(A B C)_{lmn} e^{i\Omega_{lmn}t}$,
does not always have cyclic symmetry, even if $A_{lmn}(t)$, $B_{lmn}(t)$ and $C_{lmn}(t)$
are cyclic cubic matrices.
Note that this product is, in general, neither commutative nor associative; that is,
$(ABC)_{lmn} \neq (BAC)_{lmn}$ and
$(AB(CDE))_{lmn} \neq (A(BCD)E)_{lmn} \neq ((ABC)DE)_{lmn}$.
The triple-commutator is defined by
\begin{eqnarray}
&~& [A(t), B(t), C(t)]_{lmn} \equiv (A(t)B(t)C(t) + B(t)C(t)A(t) 
+ C(t)A(t)B(t) \nonumber \\
&~& ~~~ - B(t)A(t)C(t) - A(t)C(t)B(t) - C(t)B(t)A(t))_{lmn} .
\label{T-comm}
\end{eqnarray}
The triple-anticommutator is defined by
\begin{eqnarray}
&~& \{A(t), B(t), C(t)\}_{lmn} \equiv (A(t)B(t)C(t) + B(t)C(t)A(t) 
+ C(t)A(t)B(t) \nonumber \\
&~& ~~~ + B(t)A(t)C(t) + A(t)C(t)B(t) + C(t)B(t)A(t))_{lmn} .
\label{T-anticomm}
\end{eqnarray}
If $A_{lmn}(t)$, $B_{lmn}(t)$ and $C_{lmn}(t)$ are hermitian matrices,
$i[A(t), B(t), C(t)]_{lmn}$ and $\{A(t), B(t), C(t)\}_{lmn}$ are also hermitian cubic matrices. 
The generalized bracket is defined by use of the triple-commutator (\ref{T-comm}) as
\begin{eqnarray}
{\cal{B}}(A, B, C)_{lmn} &\equiv& {1 \over i \hbar_C}[A(t), B(t), C(t)]_{lmn} ,
\label{cubic-bracket}
\end{eqnarray}
where $\hbar_C$ is a new physical constant.
By definition, we find that the generalized bracket (\ref{cubic-bracket}) has the properties of
skew-symmetry and linearity, as seen from the relations
\begin{eqnarray}
&~&[A(t), B(t), C(t)]_{lmn} = [B(t), C(t), A(t)]_{lmn} \nonumber \\
&~& ~~~~~~~~ = [C(t), A(t), B(t)]_{lmn} = - [C(t), B(t), A(t)]_{lmn} \nonumber \\
&~& ~~~~~~~~ = - [B(t), A(t), C(t)]_{lmn} = - [A(t), C(t), B(t)]_{lmn} ,
\label{tricomm-1}\\
&~&[A(t) + B(t), C(t), D(t)]_{lmn} = [A(t), C(t), D(t)]_{lmn} \nonumber \\
&~& ~~~~~~~~~~~~~~~~~~~~~~~~~~~~~~~~~~~~~~~ + [B(t), C(t), D(t)]_{lmn} .
\label{tricomm-2}
\end{eqnarray}
Note that neither the fundamental identity nor the derivation rule necessarily holds for generic variables.
(See Appendix A for properties of the triple-commutator $[A, B, C]$.)

We impose the following conditions on the canonical triplet $X_{lmn}(t)$, $Y_{lmn}(t)$ and $Z_{lmn}(t)$:
\begin{eqnarray}
&~& [X(t), Y(t), Z(t)]_{lmn} = i \hbar_C \Theta_{lmn} , \nonumber \\
&~& [X(t), X(t), Z(t)]_{lmn} = \cdots = [Z(t), Z(t), Z(t)]_{lmn} = 0 .
\label{qc-XYZ}
\end{eqnarray}
Here, $\Theta_{lmn}$ can be a normal cubic matrix, because the conditions should be imposed
time independently, and an arbitrary normal cubic matrix is a constant of motion, as seen below.

The cyclic cubic matrix $A_{lmn}(t)$ yields the generalization of 
the Heisenberg equation
\begin{eqnarray}
{d \over dt}A_{lmn}(t) = i \Omega_{lmn} A_{lmn}(t) 
= {1 \over i\hbar_C} [A(t), K, H]_{lmn} ,
\label{cH-eq}
\end{eqnarray}
where $K$ and $H$ are the Hamiltonians given by
\begin{eqnarray}
&~& K_{lmn} =  \delta_{lm} k_{mn}
+  \delta_{mn} k_{nl} + \delta_{nl} k_{lm} 
\label{K} 
\end{eqnarray}
and
\begin{eqnarray}
&~& H_{lmn} =  \delta_{lm} h_{mn}
+  \delta_{mn} h_{nl} + \delta_{nl} h_{lm} ,
\label{H} 
\end{eqnarray}
respectively.
By use of (\ref{K}) and (\ref{H}), $\Omega_{lmn}$ can be written 
\begin{eqnarray}
&~&  \Omega_{lmn} =  {1 \over \hbar_C} \Bigl(k_{ml} h_{mn} + k_{nm} h_{nl} + k_{ln} h_{lm}  \nonumber \\
&~& ~~~~~~~~~~~~~~~~~~~~~  - h_{ml} k_{mn} - h_{nm} k_{nl} - h_{ln} k_{lm}\Bigr) .
\label{Omega} 
\end{eqnarray}
Because the Hamiltonians $K_{lmn}$ and $H_{lmn}$ are normal forms, 
we find that an arbitrary normal cubic matrix $A^{(N)}$ 
is a constant of motion: $i \hbar_C dA^{(N)}/dt = [A^{(N)}, K, H]_{lmn} = 0$. 
The Hamiltonians are conserved quantities, and 
the time evolution of $A_{lmn}(t)$ is regarded as the symmetry transformation generated by them.
The fundamental identity with $K_{lmn}$ and $H_{lmn}$ must hold in order to preserve
the bracket structure under the transformation.
This requirement is equivalent to the requirement that $\Omega_{lmn}$ be a 3-cocycle, i.e.,
$(\delta \Omega)_{lmnk} = 0$.
An extreme case in which $\Omega_{lmn}$ is a 3-cocycle is that in which $k_{lm} = 1$ or $h_{lm} = 1$.
When $k_{lm} = 1$, $K_{lmn}$ and $\Omega_{lmn}$ can be written
\begin{eqnarray}
&~& K_{lmn} =  \delta_{lm} + \delta_{mn} + \delta_{nl} ,
\label{K=I} \\
&~&  \Omega_{lmn} =  {2 \over \hbar_C} \Bigl(h_{lm}^{(-)} + h_{mn}^{(-)} + h_{nl}^{(-)}\Bigr) ,
\label{Omega1} 
\end{eqnarray}
respectively.
Here $h_{lm}^{(-)} = {1 \over 2} (h_{lm} - h_{ml})$.
In this case, our generalization of cubic matrix mechanics is equivalent to 
ordinary cubic matrix mechanics
\footnote{In cubic matrix mechanics, $K_{lmn}$ is given by $I_{lmn} = \delta_{lm} (1 - \delta_{mn}) + 
\delta_{mn} (1 - \delta_{nl}) + \delta_{nl} (1 - \delta_{lm})$.
The difference between $K_{lmn}$ in (\ref{K=I}) and $I_{lmn}$ has no effect on the equations of
motion, because there is the identity $[A, B, \Delta]_{lmn} = 0$
for arbitrary cyclic cubic matrices $A$ and $B$ and $\Delta_{lmn} = \delta_{lm} \delta_{mn}$.}
discussed in Refs. \cite{YK} and \cite{YK2}.

Next, we consider the case that $k_{lm} = - k_{ml}$ and $h_{lm} = - h_{ml}$ and
both $k_{lm}$ and $h_{lm}$ are 2-cocycles, i.e., $(\delta k)_{lmn} = 0$ and $(\delta h)_{lmn} = 0$.
In this case, we can show that $\Omega_{lmn}$ is a 3-cocycle,
and then $\Omega_{lmn}$ can be rewritten as
\begin{eqnarray}
&~&  \Omega_{lmn} =  {3 \over \hbar_C} \Bigl(k_{ml} h_{mn}  - h_{ml} k_{mn}\Bigr) .
\label{Omega2} 
\end{eqnarray}
Rewriting this further, we have
\begin{eqnarray}
&~&  \Omega_{lmn} =  - {3 \over \hbar_C} \Bigl((k_{pl} h_{pm}  - h_{pl} k_{pm}) 
+ (k_{pm} h_{pn}  - h_{pm} k_{pn})
\nonumber \\
&~& ~~~~~~~~~~~~~~~~~~~~~~  + (k_{pn} h_{pl}  - h_{pn} k_{pl}) \Bigr) ,
\label{Omega3} 
\end{eqnarray}
where $p$ is arbitrary.
The relation (\ref{Omega3}) shows that $\Omega_{lmn}$ is a 3-coboundary, and
it leads to the conjecture that generalized cubic matrix mechanics can be reduced to
cubic matrix mechanics by a suitable change of Hamiltonians.
We discuss a simple example for variables that yield 
the generalized Heisenberg equation (\ref{cH-eq}) in Appendix B.

The generalized bracket structure (\ref{cubic-bracket}) is preserved by the infinitesimal 
transformation
\begin{eqnarray}
\delta A_{lmn}(t) = {1 \over i\hbar_C}[A(t), G^{(N)}_1, G^{(N)}_2]_{lmn} \delta s ,
\label{c-inf-unitary-tr}
\end{eqnarray}
where $G^{(N)}_1$ and $G^{(N)}_2$ are normal cubic matrices 
and $\widetilde{(G^{(N)}_1 G^{(N)}_2)}_{lmn}$ is a 3-cocycle.
Here, we use the fact that the fundamental identity holds for such normal cubic matrices
$G^{(N)}_1$ and $G^{(N)}_2$, so that
\begin{eqnarray}
&~& [[A, B, C], G^{(N)}_1, G^{(N)}_2]_{lmn} =  [[A, G^{(N)}_1, G^{(N)}_2], B, C]_{lmn} \nonumber \\
&~& ~~~~~~~~~~~ + [A, [B, G^{(N)}_1, G^{(N)}_2], C]_{lmn}  + [A, B, [C, G^{(N)}_1, G^{(N)}_2]]_{lmn} .
\label{C-fund}
\end{eqnarray}
Further, we find that the derivation rule
\begin{eqnarray}
&~& [ABC, G^{(N)}_1, G^{(N)}_2]_{lmn} = ([A, G^{(N)}_1, G^{(N)}_2]BC)_{lmn} \nonumber \\
&~& ~~~~~~~~~~~~~ + (A[B, G^{(N)}_1, G^{(N)}_2]C)_{lmn} + (AB[C, G^{(N)}_1, G^{(N)}_2])_{lmn} 
\label{C-D-rule}
\end{eqnarray}
holds for $G^{(N)}_1$ and $G^{(N)}_2$ if $(ABC)_{lmn}$ is a cyclic cubic matrix.

\subsection{Correspondence to Nambu mechanics}

We now discuss the relation between Nambu mechanics and generalized cubic matrix mechanics 
from the viewpoint of the correspondence principle.
First we review the relation between classical mechanics and quantum mechanics.
A physical variable $F(t)$ in CM is regarded as a linear combination of
one-index objects (a $1 \times 1$ matrix) in the form
\begin{eqnarray}
F(t) = \sum_n F_n e^{i \Omega_n t} ,
\label{Fn} \end{eqnarray}
where $F_n^* = F_{-n}$, because $F(t)$ should be a real quantity, and 
the angular frequency $\Omega_n$ is an integer multiple of the basic frequency $\omega$, 
i.e. $\Omega_n = n \omega$.
By use of the fact that the action variable $J = {1 \over 2\pi} \oint p dq$ is 
the canonical conjugate of the angle variable $\omega t$,
the equation of motion for $F(t)$ can be written
\begin{eqnarray}
{d \over dt}F(t) = \sum_n i n \omega F_{n} e^{i\Omega_n t} 
= \{F(t), H\}_{\scriptsize{\mbox{PB}}} ,
\label{HC-eq}
\end{eqnarray}
where $\{*, *\}_{\scriptsize{\mbox{PB}}}$ is the Poisson bracket with respect to
the canonical pair $\omega t$ and $J$,
and we use Hamilton's canonical equation for the angle variable,
\begin{eqnarray}
{d \over dt}(\omega t) = \{\omega t, H\}_{\scriptsize{\mbox{PB}}} = {\partial H \over \partial J} .
\label{omega-eq}
\end{eqnarray}
Under the guidance of Bohr's correspondence principle,
there is the following correspondence between $\omega$ and $\Omega_{mn}$:
\begin{eqnarray}
\omega = {\Omega_{\Delta n} \over \Delta n}  \Longleftrightarrow 
\lim_{{\Delta n \over n} \to 0} {\Omega_{n + \Delta n n} \over \Delta n}
= \lim_{{\Delta n \over n} \to 0} {E_{n + \Delta n} - E_{n} \over \hbar \Delta n} .
\label{omega}
\end{eqnarray}
Here, $\Longleftrightarrow$ indicates the correspondence,
and we use the Bohr frequency condition $\hbar \Omega_{mn} = E_m - E_n$.
We find that the equation on the right-hand side in (\ref{omega}) corresponds to (\ref{omega-eq})
with the Bohr-Sommerfeld quantization condition, $J = \hbar n$.

Next, we study the $\lq$classical' limit of generalized matrix mechanics 
based on $3$-index objects, whose frequency
condition is given by (\ref{Omega2}).
We consider the case 
that a physical variable $A(t)$ in NM is expanded as a linear combination of
one-index objects, so that
\begin{eqnarray}
A(t) = \sum_n A_n e^{i\Omega_n t} , \label{An-NM}
\end{eqnarray}
where $A_n^* = A_{-n}$ and the angular frequency $\Omega_n$ is an integer multiple 
of the basic frequency $\omega$, i.e. $\Omega_n = n \omega$.
The equation of motion for $A(t)$ is written
\begin{eqnarray}
{d \over dt}A(t) = \sum_n i n \omega A_{n} e^{i\Omega_n t} 
= \{A(t), K, H\}_{\scriptsize{\mbox{NB}}} ,
\label{HC-eq-NM}
\end{eqnarray}
where $\{*, *, *\}_{\scriptsize{\mbox{NB}}}$ is the Nambu bracket with respect to
the canonical triplet ${\cal J}_1$, $\omega t$ and ${\cal J}_2$, 
and we use $\lq\lq$Hamilton's equation" for the angle variable $\omega t$,
\begin{eqnarray}
{d \over dt}(\omega t) = \{\omega t, K, H\}_{\scriptsize{\mbox{NB}}} 
= {\partial (K, H) \over \partial ({\cal J}_2, {\cal J}_1)} .
\label{omega-eq-NM}
\end{eqnarray}
Here ${\cal J}_1$ and ${\cal J}_2$ are conserved quantities.
(See Appendix C for the $\lq\lq$Hamilton-Jacobi formalism" of NM.)
It is natural to assume the existence of the following correspondence between $\omega$ and $\Omega_{lmn}$:
\begin{eqnarray}
&~& \omega = {\Omega_{\Delta N} \over \Delta N}  \Longleftrightarrow 
\lim_{{\Delta l \over l}, {\Delta n \over n} \to 0} 
{\Omega_{l l + \Delta l n} \over \Delta l \Delta n}
\nonumber \\
&~& ~~~~~~~~~~~~~~~ = \lim_{{\Delta l \over l}, {\Delta n \over n} \to 0}
 {3(k_{l + \Delta l l} h_{n + \Delta n n} - h_{l + \Delta l l} k_{n + \Delta n n}) 
\over \hbar_C \Delta l \Delta n} ,
\nonumber \\
&~& ~~~~~~~~~~~~~~~ = - {3 \over \hbar_C} \lim_{{\Delta l \over l}, {\Delta n \over n} \to 0}
 \Bigl({k_{l + \Delta l n} - k_{l n} \over \Delta l}
 {h_{l n + \Delta n} - h_{l n} \over \Delta n} 
\nonumber \\
&~& ~~~~~~~~~~~~~~~~~~~~~~~~~~~~~~~~~~~~~~ - {h_{l + \Delta l n} - h_{l n} \over \Delta l}
 {k_{l n + \Delta n} - k_{l n} \over \Delta n}\Bigr)  .  \label{omega-NM}
\end{eqnarray}
Here, $\Delta N = \Delta l \Delta n$, $\Delta l = m -l$, $\Delta n = m - n$, 
$\Longleftrightarrow$ indicates the correspondence,
and we use the frequency condition (\ref{Omega2}) with the property that
$k_{lm}$ and $h_{lm}$ are 2-cocycles.
We find that the equation on the right-hand side in (\ref{omega-NM}) corresponds to (\ref{omega-eq-NM})
if ${\cal J}_1$ and ${\cal J}_2$ are quantized in analogy to the Bohr-Sommerfeld quantization condition.
In this way, the generalized cubic matrix mechanics can be interpreted as a $\lq$quantum' generalization
of NM.

\subsection{Conjecture on operator formalism}

We have studied the structure of generalized cubic matrix mechanics using a matrix formalism.
This mechanics has an interesting algebraic structure, but
the formalism is not practical, because it is only 
applicable to stationary systems. From experience,
it is known that in order to be of practical use, operator formalism must be capable of
handling problems in a wider class of physical systems.
By analogy to quantum mechanics, we now give a conjecture on the operator formalism of generalized
cubic matrix mechanics.
First, we make the following basic assumptions.
\begin{enumerate}
\item For a given physical system, there exists a triplet of state vectors 
$|m_1;{P}_{m_1m_2m_3} \rangle$, $|m_2;{P}_{m_1m_2m_3} \rangle$ and
$|m_3;{P}_{m_1m_2m_3} \rangle$
that depend on both the quantum numbers $m_i$ (e.g., these $m_i$ 
represent $l$, $m$ or $n$) and their ordering.
Here, the ordering is represented by a permutation (denoted by ${P}_{m_1m_2m_3}$) 
for a standard ordering (e.g., $m_1 = l, m_2 = m, m_3 = n)$.

\item For every physical observable, there is a one-to-one correspondence to a linear operator $\hat{A}$.
\end{enumerate}
Under the above assumptions, it is natural to identify the cubic matrix element $A_{lmn}$
with $\hat{A} |l;{P}_{lmn} \rangle |m;{P}_{lmn} \rangle |n;{P}_{lmn} \rangle$.
In general, the quantity $A_{m_1 m_2 m_3}$ is identified with 
$\hat{A} |m_1;{P}_{m_1m_2m_3} \rangle |m_2;{P}_{m_1m_2m_3} \rangle |m_3;{P}_{m_1m_2m_3} \rangle$.
By use of (\ref{cH-eq}),
the following equations of motion for the states are derived:
\begin{eqnarray}
&~& i \hbar_C {d \over dt} |l;{P}_{lmn} \rangle 
= \sum_{l'} |l';{P}_{l'mn} \rangle [K, H]^{(mn)}_{l'l}  , \nonumber \\
&~& i \hbar_C {d \over dt} |m;{P}_{lmn} \rangle 
= \sum_{m'} |m';{P}_{lm'n} \rangle [K, H]^{(nl)}_{m'm} , \nonumber \\
&~& i \hbar_C {d \over dt} |n;{P}_{lmn} \rangle = \sum_{n'} |n';{P}_{lmn'} \rangle [K, H]^{(lm)}_{n'n} ,
\label{S-eq}  \end{eqnarray}
where $[K, H]^{(mn)}_{l'l} \equiv K_{ml'l} H_{l'nl} - H_{ml'l} K_{l'nl}$,
and we employ the Schr\"odinger picture.
By use of relations (\ref{K}) and (\ref{H}), $[K, H]^{(mn)}_{l'l}$ can be written as
\begin{eqnarray}
[K, H]^{(mn)}_{l'l} = (k_{lm} h_{l'n} - h_{lm} k_{l'n}) \delta_{ll'} .
\label{KH-HK}
\end{eqnarray}
The equations (\ref{S-eq}) are regarded as a generalization of the Schr\"odinger equation.
The time evolution of state vectors is given by 
\begin{eqnarray}
&~& |l;{P}_{lmn} \rangle = \exp\left({i \over \hbar_C} (k_{ln} h_{lm} - h_{ln} k_{lm}) t \right)
 |l;{P}_{lmn} \rangle_0 , \nonumber \\
&~& |m;{P}_{lmn} \rangle = \exp\left({i \over \hbar_C} (k_{ml} h_{mn} - h_{ml} k_{mn}) t \right)
 |m;{P}_{lmn} \rangle_0 , \nonumber \\
&~& |n;{P}_{lmn} \rangle = \exp\left({i \over \hbar_C} (k_{nm} h_{nl} - h_{nm} k_{nl}) t \right)
 |n;{P}_{lmn} \rangle_0 ,
\label{stateslmn} 
\end{eqnarray}
where the subscript $0$ indicates that the state is that at an initial time.
In the same way, the time development of state vectors for the matrix element $A_{mln}$ is given by
\begin{eqnarray}
&~& |l;{P}_{mln} \rangle = \exp\left({i \over \hbar_C} (k_{nl} h_{lm} - h_{nl} k_{lm}) t  \right)
 |l;{P}_{mln} \rangle_0 , \nonumber \\ &~& |m;{P}_{mln} \rangle = \exp\left({i \over \hbar_C} (k_{mn} h_{ml} - h_{mn} k_{ml}) t \right)
 |m;{P}_{mln} \rangle_0 , \nonumber \\
&~& |n;{P}_{mln} \rangle = \exp\left({i \over \hbar_C} (k_{nl} h_{nm} - h_{nl} k_{nm}) t \right)
 |n;{P}_{mln} \rangle_0 .
\label{statesmln} 
\end{eqnarray}
From (\ref{stateslmn}) and (\ref{statesmln}),
we can identify $|l;{P}_{mln} \rangle$ with the complex conjugate of $|l;{P}_{lmn} \rangle$.
It is seen that this identification is consistent with the skew-symmetric property of the phase
factor in (\ref{C}).

\section{Conclusions and discussion}

We have proposed a generalization of cubic matrix mechanics
by introducing a canonical triplet and studied the structure of the mechanics
and the relation to Nambu mechanics.
The basic structure of generalized cubic matrix mechanics is summarized as follows.  
The infinitesimal symmetry transformation of a physical quantity $A_{lmn}(t)$, 
which is a cyclic cubic matrix, is given by 
$\delta A_{lmn}(t) = {1 \over i\hbar_C}[A(t), G_1, G_2]_{lmn} \delta s$.
Here the triple-commutator is the counterpart of the Nambu bracket in NM,
and $G_{1}$ and $G_{2}$ are generators of the transformation, which are normal cubic matrices.
The time evolution of $A_{lmn}(t)$ is regarded as the symmetry transformation generated by
the Hamiltonians $K_{lmn}$ and $H_{lmn}$, such that 
$i \hbar_C \delta A_{lmn}(t) = [A(t), K, H]_{lmn} \delta t$,
which is a generalization of the Heisenberg equation.
A normal cubic matrix, $G^{(N)}_{lmn}$, is a constant of motion; i.e., 
$i \hbar_C dG^{(N)}_{lmn}/dt = [G^{(N)}, K, H]_{lmn} = 0$.
The fundamental identity and the derivation rule hold 
in the case that they contain a special type of conserved quantities, $G_1$ and $G_2$, such as 
(\ref{C-fund}) and (\ref{C-D-rule}), and 
the bracket structure is preserved under the symmetry transformation, as
seen from the fundamental identity.
There is a correspondence between generalized cubic matrix mechanics and NM,
and hence our matrix mechanics can be interpreted as a $\lq$quantum' version of NM.
There is a simple system of harmonic oscillators 
described by $3 \times 3 \times 3$ matrices, which yield
the generalization of the Heisenberg equation (\ref{cH-eq}), but this is not a non-trivial
example entirely.
The dynamical variables in this system are essentially $X_{lmn}(t)$ and $Y_{lmn}(t)$, 
and the introduction of a special type of normal cubic matrices $F^i_{lmn}$ seems tricky.
Moreover, the system can also be described in terms of cubic matrix mechanics.
It would be interesting to find a non-trivial 
system, where all members of a canonical triplet are time-dependent and satisfy
(\ref{cH-eq}), and study its dynamics and relation to reality.
For this purpose, it would be useful to explore a 
counterpart to the rigid rotator that yields the Euler equation.

There still exist several obstacles that must be overcome before we can arrive at a final formulation.
For example, there is the conjecture that no global symmetry exists 
in a quantum theory including gravity.\cite{QG}
If this conjecture holds, then we would need a formulation including local symmetries.
Another modification would be necessary if we incorporate a gravitational interaction.
The theory should be formulated in a background-independent way, as the theory of general relativity.
Therefore, the scheme discussed in this paper can be interpreted as an effective description of an
underlying mechanics after fixing the background geometry and ignoring 
dynamical degrees of freedom for the graviton.

\section*{Acknowledgements}
We would like to thank Professor S. Odake for useful discussions. 

\appendix
\section{Features of Triple-Commutator}

In this appendix, we study the properties of the triple-commutator $[A, B, C]$
for cyclic cubic matrices $A_{lmn}$ $B_{lmn}$ and $C_{lmn}$.
This commutator is written
\begin{eqnarray}
[A, B, C]_{lmn} &=& A_{lmn} \widetilde{(BC)}_{lmn} + B_{lmn} \widetilde{(CA)}_{lmn} 
+ C_{lmn} \widetilde{(AB)}_{lmn} \nonumber \\
&~& ~~~~~~~~~~~~~~~~~ + ([A,B,C])^0_{lmn} ,
\label{ABC}
\end{eqnarray}
where $\widetilde{(BC)}_{lmn}$ and $([A,B,C])^0_{lmn}$ are defined by
\begin{eqnarray}
\widetilde{(BC)}_{lmn} &\equiv& B_{lnn} C_{nmn} + B_{lml} C_{lln} + B_{mmn} C_{lmm} \nonumber \\
&~& ~~~~ - B_{nmn} C_{lnn} - B_{lln} C_{lml} - B_{lmm} C_{mmn}
\label{tildeB} 
\end{eqnarray}
and 
\begin{eqnarray}
&~& ([A,B,C])^0_{lmn} \equiv \sum_{k \neq l, m, n} \Bigl( A_{lmk} (B_{lkn} C_{kmn} - C_{lkn} B_{kmn}) \nonumber \\
&~&  + B_{lmk} (C_{lkn} A_{kmn} - A_{lkn} C_{kmn}) + C_{lmk} (A_{lkn} B_{kmn} - B_{lkn} A_{kmn})\Bigr)  ,
\label{ABC0}
\end{eqnarray}
respectively.
The features of $\widetilde{(BC)}_{lmn}$ are as follows:
\begin{enumerate}
\item  $\widetilde{(BC)}_{lmn}$ possesses skew-symmetry with respect to permutations among indices:
\begin{eqnarray}
&~& \widetilde{(BC)}_{lmn} = \widetilde{(BC)}_{mnl} = \widetilde{(BC)}_{nlm} \nonumber \\
&~& ~~~~~~~~~~~~~~~~~~~ = - \widetilde{(BC)}_{nml} = - \widetilde{(BC)}_{mln} = - \widetilde{(BC)}_{lnm} .
\label{BC-cyclic}
\end{eqnarray}


\item  If $b_{lm} \equiv B_{llm}$ and $c_{lm} \equiv C_{llm}$ are 2-cocycles, i.e, 
$b_{lm} (= - b_{ml}) = b_{ln} + b_{nm}$ and $c_{lm} (= - c_{ml}) = c_{ln} + c_{nm}$,
then $\widetilde{(BC)}_{lmn}$ is a 3-cocycle: 
\begin{eqnarray}
(\delta{\widetilde{(BC)}})_{lmnk} \equiv 
\widetilde{(BC)}_{lmn} - \widetilde{(BC)}_{lmk} + \widetilde{(BC)}_{lnk} - \widetilde{(BC)}_{mnk} = 0 .
\label{B-2b}
\end{eqnarray}
\end{enumerate}


We can show the following relations from the above expressions and properties.
\begin{enumerate} \item For arbitrary cyclic cubic matrices $A$ and $B$, $[A, B, \Delta]_{lmn} = 0$
with $\Delta_{lmn} = \delta_{lm} \delta_{mn}$.

 \item For arbitrary normal cubic matrices $B^{(N)}_{lmn}$ and $C^{(N)}_{lmn}$,
the triple-commutator among $A$, $B^{(N)}_{lmn}$ and $C^{(N)}_{lmn}$ is
given by  $[A, B^{(N)}, C^{(N)}]_{lmn} = A_{lmn} \widetilde{(B^{(N)}C^{(N)})}_{lmn}$.
 \item The triple-commutator among
arbitrary normal cubic matrices $A^{(N)}_{lmn}$, $B^{(N)}_{lmn}$ and $C^{(N)}_{lmn}$
is vanishing; that is,  $[A^{(N)}, B^{(N)}, C^{(N)}]_{lmn} = 0$.

\item The fundamental identity holds if any two of $A$, $B$, $C$, $D$ and $E$ are normal forms
(e.g., $D = D^{(N)}_{lmn}$ and $E = E^{(N)}_{lmn}$, and 
$\widetilde{(D^{(N)}E^{(N)})}_{lmn}$ is a 3-cocycle):
\begin{eqnarray}
&~& [[A, B, C], D^{(N)}, E^{(N)}]_{lmn} =  [[A, D^{(N)}, E^{(N)}], B, C]_{lmn} \nonumber \\
&~& ~~ + [A, [B, D^{(N)}, E^{(N)}], C]_{lmn}  + [A, B, [C, D^{(N)}, E^{(N)}]]_{lmn} .
\label{Fund}
\end{eqnarray}

\item The derivation rule holds for normal cubic matrices $D^{(N)}_{lmn}$ and $E^{(N)}_{lmn}$,
so that
\begin{eqnarray}
&~& [ABC, D^{(N)}, E^{(N)}]_{lmn} = ([A, D^{(N)}, E^{(N)}]BC)_{lmn} \nonumber \\
&~& ~~~ + (A[B, D^{(N)}, E^{(N)}]C)_{lmn}  + (AB[C, D^{(N)}, E^{(N)}])_{lmn}  ,
\label{D-rule}
\end{eqnarray}
if $\widetilde{(D^{(N)}E^{(N)})}_{lmn}$ is a 3-cocycle
and $(ABC)_{lmn}$ is a cyclic cubic matrix.
\end{enumerate}

\section{Example}

Here we study a simple example for variables that yield
the generalization of the Heisenberg equation (\ref{cH-eq}).
The variables are three kinds of cyclic $3 \times 3 \times 3$ matrices
defined by
\begin{eqnarray}
&~& X_{lmn}(t) \equiv {\hbar_C \over \sqrt{2}} |\varepsilon_{lmn}| e^{i \Omega_{lmn}t} , 
\nonumber \\ &~& Y_{lmn}(t) \equiv {\hbar_C \over i\sqrt{2}} \varepsilon_{lmn} e^{i \Omega_{lmn}t} , 
\nonumber \\
&~& Z_{lmn} \equiv -{\hbar_C \over \sqrt{6}} \Bigl(\delta_{lm} \varepsilon_{mn} + \delta_{mn} \varepsilon_{nl}
 + \delta_{nl} \varepsilon_{lm}\Bigr) ,
\label{XYZ3*3*3}
\end{eqnarray} where each of the indices $l$, $m$ and $n$ runs from 1 to 3, 
$\varepsilon_{lmn}$ is the Levi-Civita symbol, and $\varepsilon_{lm} = \sum_k \varepsilon_{lmk}$.
The above variables satisfy the relations
\begin{eqnarray}
&~& [X(t), Y(t), Z]_{lmn} = - \hbar_C^2 W_{lmn}, ~~ [Y(t), Z, W]_{lmn} = - \hbar_C^2 X_{lmn}(t), 
\nonumber \\
&~& [Z, W, X(t)]_{lmn} = \hbar_C^2 Y_{lmn}(t), ~~ [W, X(t), Y(t)]_{lmn} = - \hbar_C^2 Z_{lmn} ,
\label{XYZ-rel}
\end{eqnarray}
where $W_{lmn}$ is proportional to $I_{lmn}$ and defined as
\begin{eqnarray}
W_{lmn} \equiv i {\hbar_C \over \sqrt{6}} \Bigl(\delta_{lm} (1 - \delta_{mn}) 
+ \delta_{mn} (1 - \delta_{nl}) + \delta_{nl} (1 - \delta_{lm})\Bigr) .
\label{W3*3*3}
\end{eqnarray} 
When we consider $X_{lmn}(t)$, $Y_{lmn}(t)$ and $Z_{lmn}$ as a canonical triplet,
the first relation in (\ref{XYZ-rel}) is regarded as the first condition in (\ref{qc-XYZ}).

Next, we introduce normal cubic matrices defined by
\begin{eqnarray}
E^i_{lmn} \equiv \delta_{lm} (\delta^i_m - \delta^i_n) 
+ \delta_{mn} (\delta^i_n - \delta^i_l) + \delta_{nl} (\delta^i_l - \delta^i_m) ,
\label{Ei}
\end{eqnarray}
where $i = 1, 2, 3$.
The quantity $e^i_{lm} \equiv \delta^i_l - \delta^i_m$ satisfies $e^i_{lm} + e^i_{mn} + e^i_{nl} = 0$, 
because $e^i_{lm}$ is a 2-coboundary.
Further, there is the relation
\begin{eqnarray}
e^i_{ml} e^j_{mn} - e^j_{ml} e^i_{mn} = - \varepsilon_{ij} \varepsilon_{lmn} .
\label{ei}
\end{eqnarray}
The variables $X_{lmn}(t)$, $Y_{lmn}(t)$ and $E^i_{lmn}$ satisfy the relations
\begin{eqnarray}
&~& \{X(t), E^i, X(t)\}_{lmn} = \{Y(t), E^i, Y(t)\}_{lmn} = \hbar_C^2 F^i_{lmn} , ~~  
\nonumber \\
&~& \{X(t), F^i, X(t)\}_{lmn} = \{Y(t), F^i, Y(t)\}_{lmn} = \hbar_C^2 E^i_{lmn} ,
\label{XYEF-rel}
\end{eqnarray}
where $F^i_{lmn}$ are normal cubic matrices given by
\begin{eqnarray}
&~& F^i_{lmn} =  
\delta_{lm} (|\varepsilon_{mn}| \delta^i_m - |\varepsilon_{mni}|) 
+ \delta_{mn} (|\varepsilon_{nl}| \delta^i_n - |\varepsilon_{nli}|) \nonumber \\
&~& ~~~~~~~~~~~~~~~~~~~~~~~ + \delta_{nl} (|\varepsilon_{lm}| \delta^i_l - |\varepsilon_{lmi}|) .
\label{Fi}
\end{eqnarray}
Here, we use the formula of triple-anticommutator
\begin{eqnarray}
\{A, B, C^{(N)}\}_{lmn} &=& A_{lmn} \widehat{(BC^{(N)})}_{lmn} + B_{lmn} \widehat{(AC^{(N)})}_{lmn}
\nonumber \\
&~& + \delta_{lm} \Bigl(\sum_k (A_{mnk} B_{nmk} + B_{mnk} A_{nmk}) c_{mk}\Bigr)
\nonumber \\
&~& ~~~ + \delta_{mn} \Bigl(\sum_k (A_{nlk} B_{lnk} + B_{nlk} A_{lnk}) c_{nk}\Bigr) 
\nonumber \\
&~& ~~~~~~~ + \delta_{nl} \Bigl(\sum_k (A_{lmk} B_{mlk} + B_{lmk} A_{mlk}) c_{lk}\Bigr) , \label{ABC-anti}
\end{eqnarray}
where $C^{(N)} = \delta_{lm} c_{mn} + \delta_{mn} c_{nl} + \delta_{nl} c_{lm}$ and
$\widehat{(BC^{(N)})}_{lmn}$ is defined by
\begin{eqnarray}
&~& \widehat{(BC^{(N)})}_{lmn} \equiv B_{lnn} c_{nm} + B_{lml} c_{ln} + B_{mmn} c_{ml}
\nonumber \\
&~& ~~~~~~~~~~~~~~~~~~~~~~~~~~~~~~~~ + B_{lmm} c_{mn} + B_{nmn} c_{nl} + B_{lln} c_{lm} .
\label{widehat}
\end{eqnarray}

When the Hamiltonians are given by
\begin{eqnarray}
K_{lmn} =  E^i_{lmn} , ~~ H_{lmn} = {1 \over 3} \hbar_C \Omega E^j_{lmn} ,
\label{K&H}
\end{eqnarray}
the frequency $\Omega_{lmn}$ can be written
\begin{eqnarray}
\Omega_{lmn} =  - \Omega \varepsilon_{ij} \varepsilon_{lmn} ,
\label{Omegalmn} 
\end{eqnarray}
by use of (\ref{Omega}).
Then, the time-development of the variables $X_{lmn}(t)$, $Y_{lmn}(t)$ and $Z_{lmn}$ are
given by
\begin{eqnarray}
&~& {d \over dt} X_{lmn}(t) = {1 \over i\hbar_C}[X(t), K, H]_{lmn} =  \Omega Y_{lmn}(t) ,
\label{H-eqX} \\
&~& {d \over dt} Y_{lmn}(t) = {1 \over i\hbar_C}[Y(t), K, H]_{lmn} = - \Omega X_{lmn}(t) ,
\label{H-eqY} \\
&~& {d \over dt} Z_{lmn} = {1 \over i\hbar_C}[Z, K, H]_{lmn} = 0 ,
\label{H-eqZ}
\end{eqnarray}
where we take $i=1 (2, 3)$ and $j=2 (3,1)$.
We find that $X_{lmn}(t)$ and $Y_{lmn}(t)$ describe a harmonic oscillator 
from the above equations.
The quantities $K$ and $H$ are expressed in terms of $X_{lmn}(t)$ and $Y_{lmn}(t)$ as\footnote{
These expressions for $K$ and $H$ are not unique.}
\begin{eqnarray}
K_{lmn} = {1 \over 2 \hbar_C^2} \Bigl(\{X(t), F^i, X(t)\}_{lmn} + \{Y(t), F^i, Y(t)\}_{lmn}\Bigr)  , 
\nonumber \\
H_{lmn} = {\Omega \over 6 \hbar_C} \Bigl(\{X(t), F^j, X(t)\}_{lmn} + \{Y(t), F^j, Y(t)\}_{lmn}\Bigr)  .
\label{K&Hansatz} 
\end{eqnarray}
It is known that equations of the same forms as (\ref{H-eqX}) and (\ref{H-eqY}) are derived
in cubic matrix mechanics\cite{YK}.  
We have
\begin{eqnarray}
&~& {d \over dt} X_{lmn}(t) = {1 \over i\hbar_C}[X(t), I, H]_{lmn} = \Omega Y_{lmn}(t) ,
\label{IH-eqX} \\
&~& {d \over dt} Y_{lmn}(t) = {1 \over i\hbar_C}[Y(t), I, H]_{lmn} = - \Omega X_{lmn}(t) , 
\label{IH-eqY}
\end{eqnarray}
where $I$ and $H$ are given by
\begin{eqnarray}
&~& I_{lmn} = \delta_{lm} (1 - \delta_{mn}) 
+ \delta_{mn} (1 - \delta_{nl}) + \delta_{nl} (1 - \delta_{lm}) , \nonumber \\
&~& H_{lmn} = {i \Omega \over 6 \hbar_C^2} [X(t), I, Y(t)]_{lmn} , \nonumber \\
&~& ~~~~~~~ = - {1 \over 6} \hbar_C \Omega  (\delta_{lm} \varepsilon_{mn} + \delta_{mn} \varepsilon_{nl} 
 + \delta_{ln} \varepsilon_{lm}) .
\label{I&H} 
\end{eqnarray}

\section{$\lq\lq$Hamilton-Jacobi Formalism" for Nambu Mechanics}

In this appendix, we study $\lq\lq$Hamilton-Jacobi formalism" for Nambu mechanics.
The basic ingredient is the differential 2-form relation\footnote{
It is known that the dynamics of relativistic strings are described by Hamilton-Jacobi formalism
based on a slightly different 2-form including two evolution parameters from (\ref{2form}).\cite{Nambu2}}
\begin{eqnarray}
d{\cal S} = x dy \wedge dz - K dH \wedge dt ,
\label{2form}
\end{eqnarray}
where $\wedge$ represents Cartan's wedge product and ${\cal S} = {\cal S}(y, z, t)$ is a differential 1-form.
Hamilton's equations (\ref{Hamilton-eq}) are derived by
taking the exterior derivatives of the above equation (\ref{2form}):
\begin{eqnarray}
0 = dx \wedge dy \wedge dz - dK \wedge dH \wedge dt .
\label{d}
\end{eqnarray}
By use of the skew-symmetric property of the Nambu bracket and the equation of motion (\ref{Hamilton-eqA}), 
we find that the Hamiltonians $K$ and $H$ are constants of motion.
The trajectory of the physical system in the phase space $(x, y, z)$ is determined by the intersection of two
surfaces, $K(x,y,z) = k =$ const and $H(x,y,z) = h =$ const.
Hereafter, we consider periodic motion on the intersection given by $C(x, y) =$ const and $z =$ const
for simplicity.

Next, we consider the canonical transformation from the canonical triplet 
$(x,y,z)$ to $({\cal J}_1, \omega t, {\cal J}_2) $,
\begin{eqnarray}
x dy \wedge dz - K dH \wedge dt = {\cal J}_1 d\theta  \wedge d {\cal J}_2 - K' dH' \wedge dt + d {\cal W} ,
\label{canonical-tr} \end{eqnarray}
where $\theta \equiv \omega t$ is the angle variable, ${\cal J}_2 = z$, and
${\cal W} = {\cal W}(y, z, {\cal J}_1, t)$ is a differential 1-form called the $\lq$generating function'.
With the relation ${\cal W} = \alpha d \beta$ and the equation (\ref{canonical-tr}),
we obtain the equations
\begin{eqnarray}
&~& x = {\partial (\alpha, \beta) \over \partial (y, z)}, ~
\theta = {\partial (\alpha, \beta) \over \partial ({\cal J}_1, {\cal J}_2)}, ~ 
{\partial (\alpha, \beta) \over \partial (y, {\cal J}_1)} = 0 ,
\label{x-eq}\\
&~& - K {\partial H \over \partial y} = {\partial (\alpha, \beta) \over \partial (y, t)}, ~
- K {\partial H \over \partial z} + K' {\partial H' \over \partial {\cal J}_2} 
= {\partial (\alpha, \beta) \over \partial (z, t)}, ~\nonumber \\
&~& K' {\partial H' \over \partial {\cal J}_1} = {\partial (\alpha, \beta) \over \partial ({\cal J}_1, t)} .
\label{KH-eq}
\end{eqnarray}
For a conserved system, we find that ${\cal J}_1$ and ${\cal J}_2$ are 
constants of motion from Hamilton's equations
for ${\cal J}_1$ and ${\cal J}_2$:
\begin{eqnarray}
{d {\cal J}_1 \over dt} = {\partial (K', H') \over \partial (\theta, {\cal J}_2)} = 0, ~~
{d {\cal J}_2 \over dt} = {\partial (K', H') \over \partial ({\cal J}_1, \theta)} = 0 .
\label{Hamilton-eqJ}
\end{eqnarray}
By solving the equations (\ref{Hamilton-eqJ}), we obtain 
the relations ${\cal J}_1 = {\cal J}_1(k, h)$ and ${\cal J}_2 = {\cal J}_2(k,h)$.

Finally, we study the change in $\theta$ over a complete cycle of $y$,
given by
\begin{eqnarray}
\Delta \theta = \oint {\partial \theta \over \partial y} dy
= \oint {\partial \over \partial y}{\partial (\alpha, \beta) \over \partial ({\cal J}_1, {\cal J}_2)} dy
= {d \over d{\cal J}_1}\oint {\partial (\alpha, \beta) \over \partial (y, {\cal J}_2)} dy
= {d \over d{\cal J}_1}\oint x dy ,
\label{rotation}
\end{eqnarray}
where we have used the equations (\ref{x-eq}). Because $\Delta \theta = 2 \pi$, ${\cal J}_1$ is given by
\begin{eqnarray}
{\cal J}_1 = {1 \over 2\pi} \oint x dy .
\label{J1}
\end{eqnarray}
Hence, ${\cal J}_1$ corresponds to the action variable in CM.
The period $T$ of a complete cycle of rotation is given by
\begin{eqnarray}
T = 2\pi {\partial ({\cal J}_2, {\cal J}_1) \over \partial (k, h)} ,
\label{T}
\end{eqnarray}
as seen from Hamilton's equation for $\theta (= {2 \pi \over T} t)$,
\begin{eqnarray}
{d \theta \over dt} = {\partial (K', H') \over \partial ({\cal J}_2, {\cal J}_1)} .
\label{Hamilton-eq-theta}
\end{eqnarray}

\end{document}